\documentclass[11pt]{article}
\usepackage{graphicx}
\usepackage{amsmath}
\usepackage{amssymb}
\usepackage{amsxtra}

\title{Influence from Future, Arguments}
\author{Holger Bech Nielsen\\The Niels 
Bohr Institute,\\ Blegdamsvej 17 - 21, 
Copenhagen {\O}; Denmark,\\ hbech@nbi.dk, 
hbechnbi@gmail.com}

\begin{document}
\maketitle

\begin{abstract}
It is the purpose of the present article 
to collect arguments for, that there should
exist in fact -although not necessarily 
yet found
- some law, 
which
imply an adjustment to special features to
occur in the future. In our own ``complex 
action model'' we suggest a version in 
which the ``goal'' according to which 
the future is being arranged is to diminsh 
the integral over time and space of the 
numerical square of the Higgs field. We 
end by suggesting that optimistically 
calculated the collected evidences by 
coincidences runs to that the chance for 
getting so good agreement by accident 
would be of the order of only 1 in 30000.
In addition we review that the 
cosmological constant being so small can
be considered evidence for some influence 
backward in time. Antropic principle may 
be considered a  way of 
simulating influence backward in time.  
\end{abstract}

\section{Introduction}
\label{s:intro}
Since long the present author and 
various collaborators\cite{old,
vacuumbomb, ownmMPP, 
ImSBled} have speculated 
on possibilities for a physical theory 
having in it some preorganization in the 
sense, that there is some law that adjust
initial condition and/or
coupling constants so as to arrange for 
special ``goals'' to occur in the future.
In works with K. Nagao\cite{NagaSI,
Nagao30,Nagao29,
Nagao36, 
Nagao37, Nagao40} we sought to 
calculate, if effects of an imaginary part 
of the action of the type of the works
with M. Nimomiya\cite{ownmMPP, ImSBled}
could be so well hidden, that such a model
would be viable. 
One could even say, that it is 
speculations
about, that future could somehow act 
back
on the present and the past. Usually 
- since Darwin and Wallace - it is 
considered 
(essentially) a fundamental law of nature,
 that this kind of back action does {\em
 not} exist.
But is that trustable? In the present 
article we shall collect arguments  for 
the opposite, namely that there truly 
IS such
a back action in time.

I would like to seek to make a kind of 
review of the
evidence for such an influence from the 
future, and
use it as an excuse for talking about 
some relatively
recent works\cite{phi2minalone}, some of 
which may not 
immediately
seem to be relevant, such as my work with 
Masao\cite{Novel,1,2}
Ninomiya on ``A Novel String Field 
Theory''\cite{1,2,Novel}.
My real motivation is to look for what 
the fine tuning
problems for the various coupling 
constants may tell
us about the fundamental laws of physics, 
which we
seek to find \cite{phi2minalone}.


In other words we could say, that I 
want  
to investigate if retrocausation is 
possible and plan to argue for that there
are inddications that it is possible. 

Although the idea of having retrocausation
is generally believed not to be true there
were at least one proposal for a theory 
of that kind proposed, nemely the theory
about electromagnetic radiation being 
radiated {\em in equal amounts backward 
and forward in time} by Wheeler and 
Feynman \cite{WF}.


The work of Feynman and Wheler avoids 
influence from the future by a discussion 
of the absorbers of the light emitted 
backward and forward, a mechanism a priori
rather different form the one we use in 
our complex action model alreadt mentioned;
but the quantum mechanics interpretation 
inspired from the Feynman-Wheeler theory,
which 
is called transactional interpretation
and is due to Cramer\cite{transactional},
is the same one as the one supported by 
our complex action theory.

The plan of this talk about the 
influence from
future will be like this :
\begin{itemize}
\item{1)}Introduction
\item{2)} Listing of arguments for 
influence from future.
\item{3)} Discussion of Timereversal 
\item{4)}Why should we NOT unite initial 
state information with
equations of motion?
\item{5)} The finiteness of String Theory 
may hide – in mine and
Ninomiyas Novel String Field Theory
\cite{1,2,Novel} - an 
influence from the
future, and that might be the reason for 
it being stringtheory.
\item{6)} Some fine-tunings as if ``God’’ 
hated the
Higgs squared field

\item{7)} Bennett’s and mine argument 
that at the time the
Cosmological Constant must already have 
hat its value, when 
densities of energy so low as the present 
were unknown/did not yet occur.


\item{8)} The Multiple Point Principle 
being successful
means influence from future.
\item{9)} If we count optimistically do 
we have
sufficient evidence for a planned universe
development?
\item{10)} At the end we conclude that one must take the possibility seriously.
\end{itemize}

\section{ Listing of Arguments}
\label{listing}
Here I should like to list a series 
of arguments for that there {\em is} 
indeed some adjustment going on to achieve
some ``goals'' we may hope to guess some 
time:
\begin{itemize}
\item{A)} Funny that many religious 
people imagine, that
there is a Governor of the world, if the 
principle
preventing such government were truly 
valid.
\item{B)} Strange that the laws about 
the initial conditions
and equations of motion behave 
differently under
the CPT-like symmetry (or under 
time-reversal)
\item{C)} Cosmological constant were
very small compared to the
energy density in the beginning; how 
could it then
be selected so small, when it had no 
significance 
at that time.(argument with D.
Bennett\cite{D13062963}). An example of 
a model living up to this by having 
indeed an influence from future in the 
effecive cosmological constant is the 
model by  Nemanja Kaloper and Antonius
 Padilla
 \cite{NemanjaKaloper}
\item{D)} Several evidences for antropic 
principle, but
mostly physicists do not like it.
(Personally I would say: The antropic 
principle is much like putting in the 
experimental fact that we humans exist 
into the theory; putting in experimental 
results can always help avoiding 
finetuning problems, so a good theory 
should be more ambitious than 
have to include  such an input.)

\item{E)} Multiple Point Principle 
(almost) successful: Higgs
mass, top Yukawa coupling, and Weak scale
relative to Planck scale.

\item{F)} Our Complex Action model with 
Higgs field square
taken to dominate gives\cite{phi2minalone}:
\begin{itemize}
\item{1)} n and p+e+antineutrino suppress
Higgs field equally much (within errors).
\item{2)} The ``knee'' cut  the 
cosmic ray spectrum
down close to the effective Higgs 
threshold.
\item{3)} Nuclear matter has low binding 
energy.
\item{4)} Higgs field in vacuum at lowest 
Higgs field square.
\item{5)} Smallness of weak 
scale/Higgs field relative to 
fundamental/Planck scale.

\end{itemize}
\item{G)}It may be very hard to make an 
ultraviolet cut off, that
does not violate locally in time a 
little bit. So an
ultraviolet meaningful theory may imply 
 influence from future?
\item{H)} General Relativity allows 
closed time-like loops...(well known to 
lead to time macines by worm holes etc.)
\item{I)} Horowich and Maldacenas\cite{} 
influence backward inside
the black hole.
\item{J)} The bad luck of SSC and the 
- though too litlle - bad luck of LHC
would follow from Higgs machines getting 
bad luck.
\item{K)}With large extra dimensions 
there appear in principle a
frame dependence of which moments are 
earlier than
which due to  the frame motion in the 
extra dimension 
directions.

\item{L)} Wheeler space time foam and 
baby universes
imply almost unavoidably influence from 
future, at least small
influences from near future.
Baby universes make effective coupling
constant depending on very far away 
influences
in e.g. Time.
\item{M)} In String theory in the 
formulation of Ninomiyas
and mine (Novel SFT) the hanging together 
of
``objects’’ to strings, or chains giving 
strings
better, is put in as an initial 
condition AND IT
LOOKS ALSO AS A FINAL STATE CONDITION!

\end{itemize}
The following arguments are even more 
theoretical speculation
arguments for influence from future:

\begin{itemize}
\item{N)} When we – e.g. Astri Kleppes 
and mine derivation of space time and 
locality etc. - seek
to derive in Random Dynamics e.g. Feynman
path integral we get the complex action 
and thus future influence from it.
And seeking to derive locality we get 
left with
effective couplings, which much like in 
baby
universe theory depends on, what goes on
averaged over all space and time.
\item{O)} Were the many e-foldings in 
inflatonal organized
in order to get a big universe 
(a miracle) ?
\end{itemize}

Somewhat estetical arguments form the 
time reversal symmetry should also be 
mentioned:
\begin{itemize}

\item{P)} The usual picture: The laws 
concerning the timedevelopment – the
equations of motion – are perfectly 
invariant under the
CPT-symmetry. Nevertheless the initial 
conditions determining the
actual solution to these equations of 
motion is chosen in a way
that makes it look more and more 
complicated as 
one progresses forward
in time! (This is the law of increasing 
entropy)
Really the mystery is not why finally 
the world ends up in a state in which 
one can say almost nothing in a simple 
way; 
but we rather should take it
that a huge number of states have same 
probability/ the heat death
state. Rather it is the mystery why it 
ever were in a state that could
be described rather simply, the state in 
early big bang times, with
high Hubble expansion rate.
\item{Q)} And even more mysterious we 
could claim: Why were the Universe
in such a special state in the beginning, 
but do not also end up in
such special and simple state?
Initial State Versus Development
Laws (equations of motion) seem not to
have the same symmetry under time reversal
(or say instead CPT)

Since Newton we have distiguished 
between initial
state information and the laws for the
timedevelopment.
Seeking the great theory beyond the 
Standard
Models our best hope to progress is to 
unite some
of the information about Nature, which we 
already
have in our litterature.
One lacking unification is the 
unification of initial
state information and the equations of 
motion.
One little – may be indicative – trouble 
is that time
reversal or better CPT symmetry is valid 
for equations of motion but NOT for the 
initial state
information!
\end{itemize}

\section{Discussion of Time 
Reversal-like 
Symmetry}

Let us look now a bit on the problem for
the usual point of view and thus the 
argument for influence from the future Q).
What are the possibilities?:

\begin{itemize}

\item{1 Possibility)} CPT symmetry could 
be the 
more
fundamental and the assymmetry w.r.t. 
time direction
of the initial state information (we know 
a lot about the
start, but the future gets more and more 
chaotic) could
be due to some sort of spontaneus break 
down, as
e.g. in mine and Ninomiyas complex action 
model: 

In principle the ``initial state 
information'' could be put in
at any time, but due to some special 
conditions in a
certain time early compared to our 
era ``the actual
solution to the equations of motion 
chosen to be
realized (by Nature)'' became mainly 
determined by
this certain era early compared our era.
This should mean that in that special era
the realized solution is arranged to obey 
some relatively simply rules, e.g. some
strongly expanding universe being the 
rule.
\item{2. possibility)} The time direction 
assymetry
might be the more fundamental and the CPT
symmetry just some effective result 
comming
out of an a priori time and even CPT
noninvariant theory. So the initial 
state CPT
noninvariance were the more fundamental
feature,  
and
the CPT symmetry for laws of nature is 
only  some 
sort of
effective or ``accidental'' symmetry
\cite{RD}. It 
is wellknown that CPT largely follows from
Lorentz invariance, so that if it were 
correct 
as I
have claimed for years, that Lorentz 
invariance
could be a low energy approximation (only 
for the
``poor physicists''), then also CPT would 
be a low energy limit symmetry.
\end{itemize}

Taking the first possibility means that 
you have in principle also the possibility 
of having some influence from the future,
so that our question as to, whether such 
influence is at all possible, gets 
answered by yes; but of course the effect 
may be essentially zero such as the 
situation 
of the ``spontaneous break down '' is
realized, 
since otherwise we should already have 
observed it so safely, that we would have 
had to believe it.

If the time dominating the fixation of 
the solution as in mine and Ninomiyas 
model becomes a certain time which is
earlier than our time - but not 
neccesarily the very first moment (if 
such one should exist) - there would 
be an opposite axis for the entropy 
running on the other side of this special 
time era(that dominantly fixes the 
solution being realized). In other words 
before the 
solution-determining time-era the entropy 
would decrease! So in that ``before 
solution dominating era '' there would 
formally be influence from the future.
Of course, if we lived in such an era, 
we would invert our time axis and still 
say, that entropy grows, except if we 
get contact theoretically or truly to 
an era 
with another entropy development axis.

If the second possibility were realized,
we should expect Lorentz non-invariant 
effects in principle. We should namely
expect CPT not to be fundamentally true,
but then Lorentz invariance could only 
with violation of other presumably good 
assumtions be exact.   

If we fundamentally did not have Lorentz 
invariance it could mean that there were 
in the ``fundamental terminology''
beyond the Lorentz invariance apperance 
perhaps some fundamental frame in which 
the physics would develop strictly 
causaly, in the sense that it would 
develop more and more chaotic (i.e.
increasing entropy) and without any 
influence from the future. But logically 
it could nevertheless be so that in some 
Lorentz frames moving relative to the 
fundamental frame there could be influence from fututure. 

\section{ Why Not Unite Initial 
Conditions and Equations of 
Motion}
In looking for a unified theory of all
physics, one often finds the idea of 
seeking to unify the various simple gauge 
subgroups of the Standard Model gauge 
group into some simple gauge group such 
e.g. $SU(5)$ or groups containing $SU(5)$
as a subgroup, such as SO(10). But since 
making progress towards finding the 
``theory of everything'' is expected to 
go via  successive unifications, one 
should also possibly imagine other types 
of unifications. Here we then ask: 
Should we
not unify initial-state conditions with 
equations of motion? This is actually 
what our already in this article suggested 
complex action model (see subsection 
\ref{complexaction}) would do. It 
predicts both (something about) the 
initial conditions and of course the 
equations of motion from the form of the 
action (as usual). In this sense one 
should really 
 guess the form of the complex 
action 
so that we can 
obtain relations between
features of the initial state conditions 
and the equations of motion. We can say
that with a Standard Model real part of 
the action taken as phenomenologically 
suggested the dimensional arguments used 
to predict that the most important part 
of the imaginary part of the action 
determining (or at least providing some  
information concerning initial conditions)
and ending with that the mass square term
for the Higgs field, are results of 
of such a unification. So in this 
sense our results from this 
Higgs-dominated imaginary part 
can be considered results of a unification
of initial state conditions and equations 
of motion.     

Also the Hawking-Hartle no-boundary 
assumption for their (and others) quantum
gravity gives information about initial
state conditions, and thus it should be 
considered a unification of initial 
conditions and equations of motion.

But now one may have general worries 
about - this kind of - unifications of 
intial conditions and equations of motion,
unless one allows for the influence from 
the future:

In fact the time reversal or the CPT-like 
symmetry leads to that the unified theory
presumably should have such a symmetry,
at least both in our complex action theory 
and in the non-boundary theory there IS 
cpt-like symmetry, except that the whole 
theory is on the manifold. Therefore it
gets very hard not to have also a final 
state condition. In fact it seems only to 
be a spontaneous breaking of the symmetry 
of this type that is likley to solve 
the phenomenological problem. But then 
there appears indeed easily some remaining 
effect of influenc form the fututre.  
    
\section{ Stringtheory, 
Regularization 
Problem, and Our Novel SFT}
Only String Theory Seems to Cope
with the Cut Off problem in Nice Way!

Presumably the best argument for 
believing, that (super)String
Theory should be the theory of 
everything(T.O.E.), is that it does NOT
HAVE THE USUAL DIVERGENCE PROBLEM.
One might wonder how string theory 
manages to avoid the
problem of divergent loops. It is well 
know that by summing
up the infinitely many loops from the 
various string states the
integrand for the loop 26-momentum obtain 
a damping factor
going with an exponential of the square 
of the loop
momentum. Thus the divergence of the 
usual type got
effectively cut off.
A related property of the lowest order 
scattering amplitudes is,
that they for large transverse momenta 
fall off even with an
exponential in the square of the 
transverse momentum.
Since String theory has gravity (almost 
unavoidably) having
such wonderfull cut off of loops behavior 
is remarkable good!

\subsection{Cut off in the Light of 
Mine 
and Ninomiyas Novel String 
Field Theory}
Let us now consider the for the   success 
of the 
string theory in coping with the 
divergences plagueing the usual quantum 
field theories so important Gaussian 
cut off of the large momenta.   

As an orientation let us look at the 
transverse momentum cut off
from the point of view of mine and 
Ninomiyas novel string field
theory:

The momentum of an open string say in 
our formalism is given
by a sum over the ``contained’’ 
``objects’’, each of which has the
variables $(J,\Pi)$, i.e. 24 momenta $J$ 
and their conjugates $\Pi$, and
the total momentum of the open string 
is proportional to the sum
of the even ``objects’’, because the 
momentum contribution from
the odd ones become zero due to their 
construction as {\em difference} of
conjugate momenta of the two even 
neighbors.
The scattering is in our SFT-model is 
simply 
exchanges of ``even
objects’’, while no true interaction 
takes place, only strings are
divided and recollected, so that the 
``even objects’’ in the initial
strings get distributed into various 
final strings.

So how does the limiting/the strong 
cutting off of the transverse momenta 
come about in the optic of our model?

Although there is a divergent number 
``objects’’ in any
string in our novel string field theory, 
these ``objects’’ are
sitting in chains with strong negative 
correlation between
the momenta of neighbors (in the chains).
So any connected piece of such a chain 
never reaches
momenta much bigger than of the order of 
one over
square root of alpha prime
$\sqrt{\alpha'}$, except for 
the momentum
assigned the total strings.
So if we only split the chains of objects 
into a few connected
pieces we cannot get any combination of 
the pieces, when
recombined to final state strings, to 
contain big amounts of
momenta compared to the alpha prime 
order of magnitude
value$\sqrt{\alpha'}$. It is this 
restriction that means, 
that we get in
Veneziano model the exponential of the 
squared
momentum falling off amplitudes.

The limitation – actually exponentially 
with the square of the
momentum in the exponent, i.e. Gaussianly 
– of the amplitude of scattering for large
transverse momenta of strings coming out 
of collisions of
strings in our novel string field theory
(SFT) is due to the very
strong anti-correlation of the momenta 
of the ``objects’’ -
crudely functioning as constituents of 
the strings – so that only
very limited momenta are statistically 
found on connected
pieces of object-chains.
Since this so important - for the 
momentum cut off  –
(anti)correlation of the ``objects’’ on 
the chains used for strings
is put in as INITIAL and even as FINAL 
STATE conditions in
order to describe the strings by means 
of ``object’’-chains, one
can say that in mine and Ninomiyas SFT we 
have arranged
the transverse momentum cut off 
effectively by the initial or
final states having been assumed to have 
the appropriate (anti)correlations!

\subsection{The Limitation of 
Momenta in Loops}
For each limited loop order corresponding 
in our novel SFT to
splitting the ``cyclically ordered 
chains’’ of ``objects’’ into a limited 
number of subchains before being 
recollected into new  ``cyclically ordered 
chains’’ forming the final state strings 
(depending on the order (of loops)) the amount
of momentum, that can be sent out as 
transverse momentum in
a scattering is limited due to the 
correlations among the
``objects’’ (neighboring on the chains).
The higher the order though the higher 
is the effectively allowed order of 
magnitude of the transverse momentum,
corresponding to the wellknown faact that 
higher and higher loop order in unitarity 
corrections to the Veneziano model has 
a slower and slower fall off for large 
momenta the higher the order (i.e. the 
larger the number of loops).  
Roughly this relevant correlation 
corresponds to the ``stringness''
in the sense, that it is also this 
correlation (between neighboring ``objects''), that ensures that very
small pieces of strings carry only very 
little momentum.
But have in mind, that in OUR theory the 
hanging together to
strings is only put in as initial state (and even final state)
conditions.
Even the alpha prime $\alpha'$ scale – so 
needed to 
make a chance of
having a cut off effectively – is in our 
model {\em only put in as an initial and 
final 
state condition} (nothing in the 
completely trivial and basically 
nonexisting dynamics talks about alpha 
prime!) 

So one really in mine and Ninomiyas 
novel string field theory must ask:
String theory cut off, from where
does it come?

Generally: When one interacts (locally) 
with the string
– in our formalism or in other ones – you 
can only
transfer little – meaning given by apha 
prime
(inverse square root 
$\frac{1}{\sqrt{\alpha'}}$) – momentum 
into the
scattering.
Via Heisenberg uncertainty this is turned 
into
an extension of the strings due to 
quantum fluctuations.
But it is crucial for the effective cut 
off, that the string
hangs piecewise together; if e.g. in mine 
and
Ninomiyas novel SFT you could split 
the ``objects’’ in
a way, in which no ``objects’’ kept 
attached to their
neighbors almost, then the momentum in the
scattering could be much larger, and very 
likely a
divergence problem would reappear.

In fact it is well known that the higher 
loops one
consider in string theory (unitarity corrections to
Veneziano model) the slower becomes the
coefficient in the Gaussian fall off of 
the amplitude with the exponential of 
the square of
the transverse momenta.
This means that the more pieces the 
string – or
in our model the to the strings 
corresponding
``cyclically ordered chains'' – are cut 
into and
recollected under the scattering, the 
larger can
the transverse momentum become.

If one would attempt to split up the 
string to be
actually built form discretized elements, 
one
would be back in quantum field theory and 
it
would be as hard as usual to avoid
divergencies.
The continuity of the string – or in our 
novel
SFT formulation the cyclically ordered 
chains –
is crucial for the achievements w.r.t. 
avoiding
divergencies and keep tranverse momenta 
low.

\subsection{Looking for a Cut Off 
Machinery}
Let us now look whereto we are led when
we look for a way to make a cut off:


Now I would like to speculate as to where 
we are
led to think, if we which to get sense 
out of a theory,
in e.g. too many dimensions, so that 
ultraviolet cut
off is truly a necessity:

First we could think of  modifying 
geometry or we may 
seek to
keep it:
\begin{itemize}
\item{1)} Cut offs like lattices which 
have a discretized
geometry.
\item{2)} Keep e.g. flat geometry or at 
least a manifold.
\end{itemize}

In this second case
where are we led, if we seek a cut off of 
the untraviolet
divergencies, but cling to continuous 
manifold or let us
for simplicity say simple Minkowskian geometry (but
continuous space and time) ?

If we use point particles with 
interactions we have no
chance to get any form factors to rescue 
us against the
ultraviolet divergencies.(we might though 
use hihger order derivative on the fields in the Lagrangian density, but let us leave that as another possibility).
So we are let in the direction, that we 
must take the
particles, with which we want to work, to 
be composite
objects / bound states or rather most 
importantly
extended objects, so that interactions 
with the various
components have the chance to cancel out 
couplings to
very high momentum states (which is what 
cause the
divergencies).

Thus let us at least look  towards seeking 
 cut off in direction of
bound states:

Let us now think along the line, that we 
replace
the particles, we consider 
phenomenologically, by
bound states or composite structures. 
That is to
say, that looking more deep inside they 
shall turn
out to consist of some ``smaller’’ parts
``partons’’ say.
It is fine that we may then get form 
factors, since
they have the chance to cut off the loop
integrals and make them converge.

Now we may talk the language of Bjorken 
$x$ being
the fraction of longitudinal momentum 
carried by
a ``parton’’.

If the partons have non-zero Bjorken 
$x$, then you get
parton parton scatterings, when the 
bound states
collide and the situation is much like, 
if the partons
really existed  and we are back to the 
point particle
play: there will finally result 
divergencies again.

So if we are looking for avoiding 
divergencies we are driven
in the direction of taken all the Bjorken 
$x=0$.
But that then in succession means that 
collision of
only a few partons from one 
particle(=bound state)
with partons in the colliding 
particle(=bound state)
will hardly give any momentum transfer, 
hardly
mean even a scattering.

Once assuming $x=0$ for all the partons we will get negligible
momentum transfer by just scattering a 
few partons with
each other; that is too much cutting off.
The effective way to get some significant 
scattering to
identify with the scattering of the 
particles(=bound states),
we want phenomenologically, is to 
exchanges from one
bound state to another one  a large 
number(infinitely many)
partons.
This means we are driven towards a 
picture, in which a
scattering is mainly an exchange of some 
part of one 
 composite particle with part of 
another one.
But none of the constituents (=partons) 
truly interact.
Rather the constituents individually just 
continue
undisturbed as if not interacting at all!

Remark how we got driven towards the 
picture
of String Theory in mine and Ninomiyas 
novel
string field theory: The bound state, we 
consider
should be composed from constituents not
interacting at all!

These consituents or partons, we are driven
towards, are of course to be identified 
with the
``objects’’ in Ninomiyas and mine 
novel SFT(= string field theory);
precisely these ``objects’’ of our 
theory do not
change at all.

So we for the moment think of 
``Even Objects’’ as Partons:

Does it matter whether we consider
our ``Objects’’ as constituents or
the true string interpretation
definition of the ``Objects’’ J from
discretizing right and left movers 
in the string? For  this true definition
of the ``objects'' we have to refer to 
the other article in the present Bled 
Conference proceedings 2014 on what comes 
beyond the standard models \cite{Novel, }.
 
Very shortly let us though on the 
definition of the ``Objects’’ say:

Since the ``objects’’ are defined as the
difference between the values of say the 
right
mover component 
of $X^{\mu}(\sigma, \tau)
= X^{\mu}_R(\tau - \sigma) + X^{\mu}_L(\tau+\sigma)$ i.e. as $J^{\mu}_{RI} = X^{\mu}_R(
\tau_R(I+1/2)) - X^{\mu}_R(\tau_R(I-1/2))$
(where $\tau_R = \tau - \sigma$, and 
we imagine a discretization replacing 
$\tau_R$ by an integer number $I$ instead
and let $\tau_R(I\pm 1/2  )$ denote the 
neighboring $\tau_R$ points around the 
point corresponding to $I$ in the 
discretization.)  at two near to each 
other
values of the ONE relevant variable, it 
is in fact
proportional to the derivative of the 
right mover
component.
To reconstruct the position field we 
both have to
integrate (or sum) up and we need both 
left and
right.
On open strings boundary conditions causes
the left and right mover to be the same. 
But for
open strings they are different.

After we have identified the right with 
left mover ``objects’’
for the open string (as the boundary 
condition for open
string leads to) the objects describing an open string sits
topologically in a circle, called by us
``a cyclically ordered
chain of objects’’.
So the topology of the structure 
describing the open
string by us is a circle and and not as 
the open string
itself an interval.
But the momenta of the open string is 
written as a sum
over contributions from the ``objects’’ 
sitting along the
cyclically ordered chain (the circle).
So as long as one can consider a 
distribution of momenta to 
the various ``objects’’, we
can consider the ``objects’’ constituents
(for that momentum distribution purpose at least).

So we might ask: Can we forget the 
string and only
think on Our ``Objects’’ ?

If you go over to considering the 
``objects’’ of
our model as constituents of the composite
particle(described as the string), you 
ignore the
string as not being the right way of 
thinking of the
same theory.

Contrary to the string point of view, in 
which the
string moves internally as it moves 
along, the
``objects’’ are stale and just do not 
change (Well, their position is a bit 
more tricky to consider, so we may think 
of them as free partons).
The ``objects’’ fit with the constituents 
not
interacting but just being exchanged 
en block
from bound state to bound state.
Pieces of String Time Track per Pair
of ``Objects’’ with Lightlike Sides
Time Track of String from Pieces per
Pair of ``Objects’’ Lightlike Sides
The Very Scattering Moment, Only
Exchange of Pieces

Whatever the string may develop 
mechanically after a
collision it is an almost pure exchange 
of parts that take
place at the very collision.
At least if the hit is only at ONE POINT 
of the hitting strings,
then from locality nothing can happen at 
other places in the
very first moment.
So in the limit of infinitely many 
constituents (like continuum
string) the first moment of a scattering 
ONLY an exchange
of pieces can matter.
So, if indeed no parton with$ x$
 different from 0 is allowed in
order to make a good cut off bound state 
theory, then when
first partons hit we can ONLY have 
exchange of pieces
interaction: So in this first moment 
there is in this sense no
true scattering! (Like in mine and 
Ninomias model).

But there is a 
need for exchange of pieces

If we have $x=0$ bound states, there would
without exchange of pieces be no 
scattering, no
essential momentum transfer at all.

Now I say: We are driven – in seeking for a cut
off – to a theory with a system of 
particles
(corresponding to the strings in string 
theory)
being bound states with all partons having
Bjorken $x=0$, and they scatter only by 
exchange
of pieces. So it is essentially only how 
one
thinks the constituents as distributed 
between
the particles, that change in the 
scattering.

It is wellknown that the higher dimensions
spacetime has the more severe are the ultraviolet divergencies:
High dimensions give ultraviolet
divergences.
\subsection{
Rescuing the Species Doubler
Problem by Pushing 
Chiral Charge
to Central
Station in Extra Dimension}

In the Standard Model one has a remarkably 
tricky
cancellation of the chiral anomalies 
associated with the
(chirally coupled) gauge fields.
Non of the fermions in Standard model 
have their
``species doubler’’ (with opposite 
handedness, but same
charge combination).
So it should after mine and Ninomiyas 
no-go theorem
be impossible to put the Standard Model 
on a lattice, or
for that matter regularize it in gauge 
invariant way at all.
I.e. No cut off should exist, which can 
keep gauge
invariance.
The way Norma Mankoc Borstnik
and I attempted to escape this problem
were the following:

The way we attempted to escape the no-go
theorem was by having infinitely large 
extra
dimensions allowing superfluous fermions 
to be
pushed out to infinity.

Let me look at the nogo theorem problem by
thinking of the anomaly telling that the 
chiral
charge is not conserved, but has a lack of
conservation correction proportional 
$F \tilde{F}$(with some gauge fields put 
in for the two $F$'s).

\subsection{
Anomaly way of Looking at No-Go
Anomaly Requires Pushing out or
Fetching in Chiral Fermions}

Because of the anomaly we need locally in 
space-time to
be able to obtain extra chiral fermions 
in spite of them
having conservation laws making that 
impossible in the
regularized theory.
In Norma Mankoc Borstnik’s and mine 
attempt to cope
with Wittens no go theorem\cite{} we 
propose to 
have
non-compact extra dimensions:
Then the superfluous or missing chiral 
fermions may be
pushed out or be brought in from the 
infinitely far away in the
extra dimensions.
You almost bring them out to a mysterious 
central station
for pushed out chiral fermions, from 
where they may
reappear in the practical world later or 
earlier or
somewhere else than from where they were 
pushed out.

With such central station whereto chiral
particles are broght in and out to various 
places or times in the 3 +1 dimensional 
world is to be imagined in the model 
needed (say Norma Mankoc's and mine), then
one may suspect that one easily  get 
times mixed up having such an 
exchange station for chiral
fermions. There namely has to be somehow 
a controle that the total number of chiral
fermions of a certain type is conserved 
in the regularized model. But then how 
to get the information of the creation 
seemingly of one at a certain point in the 
3+1 space time transfered and brought 
together with the uses or further 
creations arround space time without 
endangering the no influence from future 
principle(which we attempt to attack in this article)?

If really the chiral fermions are 
fundamentally
conserved in the regularization scheme – 
here
thought upon as the true theory – but 
just seem
not to be because they are pushed out to 
an in
the extra dimensions infinitely far away 
place, it
may seem difficult to keep truly no 
influence from
future from the practical 3+1 dimensional 
point of
view.
Would one really could have the number of 
chiral
fermions being added to the central 
station for
such fermions pushed out be kept to netto 
zero
without some influence back from the 
future?

\section{Some Potential Killings of Our
Complex Action Turned Out
Supporting It.}\label{complexaction}
Funnily enough I have found a few  
cases, where seemingly arguments against 
the validity of the complex action model 
with its influence from future, actually 
get turned arround and leads to evidence 
for the influence from future instead, 
because they turn out rather to show 
that nature has just some number just 
finetuned almost to solve the problem.  
\subsection{Short Review of Complex 
Action Model}
Let me here review a bit the main 
point of the theory of the complex action.
A priori it would seem obvious that if 
we took the action $S[history]$ to be 
complex rather than as assumed in the 
usually believed theory, then one would 
immediately see that effects of 
 non-unitarity and if one used classical
calculation one would also expect that 
otherwise real variables would run complex.
In other words at first it would look as 
if the idea of taking the action complex 
is phenomenologically so bad that any 
hope is out unless the imaginary part is 
extremely small; and so if real and imaginary were about equal in size as one would 
guess there seems at first to be no chance.
But that is according to the calculations or estimations on which we are still 
working not true! Most convincingly this 
is seen in a Hamiltonian formalism, in 
which not so surprisingly a complex 
action would lead to a non-Hermiten 
Hamiltonian. In fact the main point is
that as long time has past since the start,
almost certainly the universe developping 
by the now assumed non-Hermitean 
Hamiltonian gets increasing probability 
for being in those states, which have 
the largest (eigen)values for the 
antiHermitean part (divided by i) $H_I$ 
of the 
Hamiltonian, if we think of having split 
it
as $H=H_R +iH_I$ where then $H_I = 
\frac{1}{2i}(H -H^{\dagger})$. If we now 
have assumed - as we have to assume to 
avoid 
that the Wentzel-Dirac-Feynmann-path
integral shall not be divergent due to 
the imaginary part of the action 
$S_I[history]$ going to plus infinity -
that there is an upper bound on the 
antiHermitean part $H_I$ or almost 
equevalently a lower bound on the 
imaginary part of the
action $S_I$, then we argue that the 
system after long time will arrive to 
a superposition of states with their 
(eigen)value for $H_I$ close to the 
assumed upper bound. Once we have 
argued the sytem to be in such a state 
we have the suggestive approximation
of $H_I \approx \hbox{``upper bound''}$
and can consider the antiHermitean part 
$H_I$ an approximate c-number and by a 
timedependent normalization we can 
completly remove the effect of this 
antiHermitean part. This crude argument
thus allows us to suppose that after all
the antiHermiteamn part $H_I$ of the 
Hamiltonian is not important provided 
we study what happens in a universe, that
is already very old compared to some 
fundamental scale for the theory provided
we have just an upper bound on this 
antiHermitean part. This may not be 
totally convincing as written, but we 
have formal formulations and it is 
essentially correct but in order not 
to have troubles with the Born rule 
of quantum mechanics that one shall the 
probability for measuring a state by 
using the nummerical square of the 
coefficent to a normalized states one 
shall a new inner product which we call
$|_Q$ (so that we can write $<b|_Qa>$)
with the property that w.r.t. this 
inner product the Hamiltonian $H$ gets 
{\em normal}. Normallity means that
the antiHermitean part commutes with 
the Hermitean part i.e. $[H_R,H_I]=0$.
(The $Q$ that occurs as an index to the 
new inner product $|_Q$ to be used 
instead of the original inner product 
$|$ is an operator constructed from the 
Hamiltonian - using it diagonalization - 
and then we defined 
$<a|_Qb> = <a|Q|b>$.)

Even though now we have argued, that one 
will 
obtain a timedevelopment as if there 
existed a Hermitean Hamiltonian even, when
the true Hamiltonian is not Hermitean, 
provided one uses the modified 
inner product $|_Q$, there is one very 
interesting and important effect of the 
antiHermitean part $H_I$ or of the 
imaginary 
part $S_I[history]$ of the action left:
These antiHermitean or imaginary parts 
determine the initial condition 
effectively seen! We saw already just 
above that the antiHermitean part of the 
Hamiltonian were important for the states 
into which the likelyhood of finding the 
world got larger and larger as time 
went on. So effectively in a late 
stage of the development of the 
universe it becomes most likely to 
find that this universe is in a state 
with a high -i.e. close to the upper 
bound - value for the (eigen)value
of the antiHermitean part $H_I$. 
This really means that we shall look at the complex action theory as a model
{\em unifying the initial conditions
 with the equations of motion}.

Such a unification of course is in 
principle very wellcome, if one can 
find it. In the Hamiltonian formalism 
with a non-Hermitean Hamiltonian one can
see that unless one puts the system/world
in a state that has absolutely zero 
component after some eigenvectors of the
Hamiltonian, it will go so that as
time goes on the various eigenstates in an
expansion of the actual state will grow 
up exponentially with coefficients 
going as $-i t\lambda_i$ where $\lambda_i$
is the for the coefficient relevant 
eigenvalue of the non-Hermitean 
Hamiltonian $H=H_R +iH_I$. Have in mind 
that for non-Hermitean Hamiltonian of 
course the eigenvalues $\lambda_i$ are 
typically complex. It is of course the 
imaginary part of $\lambda_i$ which gives 
rise to the time development of the 
numerical value of a coefficient 
$c_i\exp{-t\lambda_i}$ to some eigen
vector $|\lambda_i>$ (even though these
eigenvectors are not orthogonal to each 
other, one could 
still imagine using them in expansion).
Exponentially soon a rather small 
collection of the eigenstates with the 
largest - in the sense of most 
positive - imaginary parts of their 
$\lambda_i$'s will soon take over.
Thereby a rather specific development 
of the universe gets selected out and 
one can understand that the antiHermitean 
part of the Hamiltonian can have strong 
influence on which states one at a late 
stage in time is likely to find such a 
universe with non-Hermitean Halmiltonian.
Thus it is understandable that there can 
be something in the statement that the 
theory unites initial condition theory 
with equation of motion theory.

Our studies have led to that one may 
distinguish reasonably defendable ways 
of extracting the information from 
a quantum theory with a given action 
- two different ways especially 
suggestive in the case complex action -
namely 1)``with future'' and 2)``without
future''.

\subsection{Guessing the Standard 
Model
Imaginary Part of the Action}

At the present conditions in the Universe
- but not at all applicable perhaps in
the early times just after a possible big 
bang say less than $10^{-12}$ s say -
the Standard Model seems to work perfectly 
except perhaps in very high energy 
accelerators and in cosmic radiaton. So 
we should
expect that at least the real part 
$S_R[history]$ of the 
action $S[history]$ should be given well
by the action of Standard Model. Now the 
very natural guess is, that you get the 
full complex action by just letting all
the coefficients of the various terms in
the Standard Model action become complex.
You might even as the a priori most 
promissing guess think, that the phases 
are 
rather random and of order unity, meaning 
of the order of $100^0$, except though, 
that the mass term for the Higgs particle 
deserves special discussion. 

Let us remind about the discussion 
around the hierarchy or the scale problem
for the usual real action Standard Model:

If you imagine a cut off at the Planck 
scale or some new physics at some GUT 
scale at almost Planck energy scale, then 
one has the problem that corrections
to the bare Higgs mass square as written 
in the Lagrangian density $m_{Hbare}^2$ in 
order to obtain from that the measured 
mass square $m_{Hren}^2$ becomes typically
very large, either it is divergent or by 
meaans of fixing some unified scale it 
becomes when renormalized to that scale 
anyway huge compared to the scale of 
measured Higgs mass square or the weak 
scale. So it is a wellknown finetunig 
problem how to get the weak scale be small
compared to the huge scales involved in 
the loop calculations even if one renormalizes to some unifying scale. You might 
keep the corrections smaller by having supersymmetric partners - but the LHC results 
so far rather show the surprise that such 
oes are so far not found -. But whatever
might be the solution to this problem of 
how the weak scale became so small say 
compared to the Planck scale and how to 
keep it there it might it easily becomes 
so that the bare mass square $m_{Hbare}^2$ 
becomes appreciably bigger than the renormalized one $m_{Hren}^2$ numrically.In the
case when some supersymmetric particles 
exist and makes the mass square correction
{\em only} logarithmically (divergent)
the size of the bare divided by 
renormalize will though only be ``logarithmic'', which means not so phantastically 
big after all. But if the supersymmetric partners do not exist or are very heavy 
then again the bare mass square will 
typically be much larger than the 
renormalized/observed Higgs mass square.

When we now want to guess the size of the 
imaginary part of the Higgs mass square,
the suggested guess is that it should be 
of the same order as the real one; but 
now should it be as the real renormalized 
or as the real bare ? Most likely the loop
corrections for the real and for the 
imaginary parts are completely different
and huge, so the question becomes: Would 
the same mysterious fine tuning, which 
made 
the real part $m_{Hren}^2|_R =
\hbox{observed/effective Higgs mass square}$ of the renormalized 
mass square for the Higgs also function
for the imaginary part, so that in some 
way - which we may or may not understand 
- the 
effective/renormalized (whatever that 
might exactly mean) imaginary part 
of the Higgs mass square $m_{Hren}^2|_I$
becomes as small as the real 
renormalized part order of magnitudewise?

Very likely the solution to the finetuning
problem (= the scale problem) of why the 
weak scale is so low compared to the 
Planck scale say will be solved in a way 
that will not make also the 
``renormalized'' scale for the imaginary 
part of the ``Higgsmass square'' small 
compared to say the Plack scale. For 
instance this is the case for our own 
``solution'' to this problem by means 
of the multiple point principle: This 
``solution'' means, that, if we make the 
very strong assumption  that there is 
some finetuning fixing the 
parameters/coupling constants of the 
theory working in nature in a way 
restricted so that there becomes 
{\em several different vacua all having 
very small energy densities(=dark energies
= cosmological constants)} (for puposes 
of the weak scale we just say exactly zero
energy densities are assumed in the vacua) we how found a viable picture with 
strongly bound states of 6 top + 6 
anti-top quarks and a set of three 
different vacua in the Standard Model,
in which this requirement leads to an 
exponentially small value of the weak 
scale compared the scale of the Higgs 
field in one of the vacua considered 
degenrate. In other words with our 
assumtion of vacua with zero energy 
density 
(called ``multiple point principle'' 
(=MPP)) and 
some in principle calculable speculation
about bound stattes of quarks and anti 
quarks the parameters of the standard 
model need to to take such values 
that the renormalized Higgs mass square 
must be very small compared to the scale
for the Higgs field in one of the by us 
assumed vacua. We then add as an extra 
assumption to our multiple point
principle that for one of the vacua 
the Higgs field present should be of 
the order of the Planck energy. This 
latter assumption is already supported
by the parameters of the Standard Model
if one assumes this Standard Model to
be valid up to so high energies (or 
Higgs fields). It found a support together
with the muultiple point principle by
the Higgs mass found in Nature agreeing
with our PREdiction.

But really in our complex action model
physics comming out of the real and of
the imaginary part of the action are 
{\em quite different}, crudely the real 
part givesequation of motion and the 
imaginary te initial conditions, so to
expect that some mysterious mechanism 
make the same finetunin on both is not 
at all likely. Therefore we shall 
conclude that it is most likely that
there is no finetuning going on to
make the effectively 
observed/``renormalized'' imaginary part 
of the Higgs mass square small compared to
say the Planck scale value. If so, then
we should expect it to be probably of 
the order of the Planck scale. Putting 
into Standard Model extended to have 
complex action this size of the Higgs mass
square imaginary part would mean that 
considering a process of dayly life or of 
LHC the Higgs mass square term would 
give contribution to the imaginary part
of the action, which are larger than the 
contributions from the other terms by
a factor $M_{Pl}^2/(~TeV^2)\approx 10^{34}$. This means
that we  from dimensional arguments  think we 
could argue
that the most important term in the 
imaginary part of
the action should be the part from 
the Higgs mass
(square) term.

With this we argued that we under present
conditions can 
approximate the imaginary part 
of the action $S_I[history]$ by only
the contribution from the 
Higgs-mass-square term
\begin{equation}
S_I[history] \approx \int m_{Hbare}^2|_I 
|\phi_H(x)|^2 \sqrt{g} d^4x \label{SI}
\end{equation}  
(the $\sqrt{g}$ is just 4-volume measure
inserted to make the formula o.k. in the
general relativity case, but really you 
may use flat space approximation and ignore it). The Higgs field were denoted 
$\phi_H(x) $ and depends of course on 
the event coordinate (set)$x=\{ x^{\mu}\}$.
The integral is, provided we use the 
``with future''-interpretation of the 
complex action theory, to be interated 
over all space time including {\em both
future and past}, and then it is this 
quantity (\ref{SI}) which at least in 
first approximation selects intial 
conditions or what really happens by
letting the true happening  $history$ 
have the 
minimal value for the imaginary part of 
the action $S_I[history]$ among all the
say by equations of motion allowed 
possible histories. For a crude 
understanding of our complex action 
theory one may take it that it predicts 
roughly that 
\begin{equation}
S_I[\hbox{$true$ $history$}] 
\stackrel{<}{=} 
S_I[\hbox{any other 
$history$}]. \label{Godwill} 
\end{equation}
(more detailed calculations of some 
predictions may be found in \cite{Nagao36, 
Nagao37, Nagao40} and in some of the 
papers with Ninomiya \cite{own}). 

One way of putting forward the idea of 
the universe initial conditions being 
arranged in a way governed so as to 
achieve say small (or preferably 
numerically large negative) contributions 
to $S_I[history]$ is to call it a ``God''
(it is only a god in quotes(thanks to 
Mette H{\o}st)) governing the world so 
as to seek to minmize the imaginary part 
of the action $S_I[history]$. In this 
language our expression (\ref{IS}) means 
that this ``God'' only cares for the 
integral over space time of the Higgs 
field; ``He''to day  mainly care for Higgs 
particles and modifications in the Higgs 
field.  Oscillations in the Higgs field 
– meaning physical
Higgs particles – will obviously make 
the square of the
Higgs field integrated over all space 
time bigger.
So producing Higgses should e.g. be hated 
and
avoided by the ``God’’.
(Had the sign been so that it 
corresponded to ``God'' loving Higgs 
bosons instead ``He'' would have filled 
more up with Higgsbosons, say an expectation value of the Planck order of magnitude 
at least).

But if  ``He’’ hates the Higgs ``He'' 
should love the particles suppressing in
there neighborhood the Higgs field? And 
fill the whole
Universe with the most favoured ones.

It is for instance the quarks and the 
charged leptons that are surrounded by
a Yukawa potential region in which the 
Higgs field has an additional Higgs field
- the Yukawa potential -, and so a more 
strong field the bigger the mass or 
the lepton causing this field. One may 
easily understand that the Higgs field 
having in vacuum its wellknown expectation
value $<\phi_H(x)> = 246 \;  GeV$ is a bit
deminished numerically inthe 
Yukawa-potential-region around a quark
or a (charged) lepton.  Now in principle 
we do not know whether the square of the 
Higgs field $|\phi_H(x)|^2$ increases or 
decreases as one enforce a little region 
in space(-time) to have a given Higgs 
field dimished say w.r.t. the usual 
vacuum Higgs field. Intitively one would
think the square would decrease when the Higgs field itself decreases but there 
could - and indeed there are - be effects
causing it to go oppositely(as have 
argued for below and in
the articles\cite{phi2minalone}).
In any case unless there is just an 
extremum of the square $ <|\phi_H(x)|^2>$
as a function of the Higgs field itself
$<\phi_H(x)>$ in the usual vacuum 
situation there would be an effect 
positive or negative upon the imaginary
action $S_I[history]$ as given by 
(\ref{SI}) from the Yukawa-poteential 
regions around the quarks or 
(charged)leptons, because the normal 
Higgs fields a bit suppressed in such
Yukawa field neighborhoods.

This means that e.g. the ``God'' would 
either love or hate these quarks and 
charged leptons, and that the more 
strongly the heavier they and the stronger
they therefore couple to the Higgsfield.

This in turn means that e.g. a particle 
like the neutron with its three valence 
quarks and further quark pairs inside it
will suppress the Higgs field from its
usual vacuum value a bit and then 
depending on the sign of the derivative
$\frac{d<|phi_H(x)|^2>}{d<\phi_H(x)>}$ 
increase or decrease the imaginary part 
of the action $S_I[history]$, thus 
the neutron would be respectively 
hated or loved by ``God''.

Now in nature one can by weak interactions
get a neutron transformed into a proton,
an electron and an electron-anti-neutrino.
Thus if the ``God'' loved say the netron 
itself more than the proton the electron 
and the electron-anti-neutrino together
we would expect that ``He'' would 
have arranged initial conditions 
- and if ``He'' were allwoed to it 
also that coupling constants or whatever
could help - so as to make there be only 
neutrons but no protons and elctrons etc.
We know from astronomy and our own 
earth neighborhood that there exist both
neutrons and protons and electrons 
(and even neutrinoes) in rather large 
amounts, none of them being truly
so much suppressed compared to the other.

{\bf At first we may look at this 
fact there there are both neutrons and 
protons in the world today as a 
falsification of the minmization of 
imaginary part of action ideas!} 

It becomes in our complex action theory 
an embarrasing question:
Why not only n or only
p+e+antineutrino ?

An idea to an attempt to disprove our 
complex
action model with the Higgs field square
integrated as the imaginary part of the 
action:
Why do we not have either?:
\begin{itemize}
\item{1)} Only neutrons n and no protons 
nor
electrons, or
\item{2)} Only protons with their elctrons e and
antineutrinos, but no neutrons at all.
\end{itemize}

Either one or the other would probably be
favoured and thus by ``God’’ be arranged 
to be
realized!

\subsection{Solution to: Why both 
protons
and neutrons?}
Actually this problem of why not 
only protons( with their electrons) or 
only neutrons in the world in our complex 
action model has the ``solution'':

{\em If the neutron is exactly equally 
much ``loved'' as the the proton the 
electron and the electron-anti-neutrino
together -in the sense of contributing 
the same to the imaginary part of action
$S_I[history]$, then there would be no
reason for ``God'' to eradicate one of 
the two types of particles. But this
requires a certain relation between 
the masses of the quarks corrected 
by their Lorentz contraction factors
and the electron mass. But remarkably
this relation is satisfied within 
calculational accuracy!} (light 
quark masses are rather badly known 
so the accuracy is not so high)   

Basically\cite{phi2minalone} in order 
that there shall be no reason to either 
remove from the world the neutrons nor 
the combinations of protons and electrons
(the neutrinoes anyhow contribute much
less to the imaginary part than the 
massive quarks or leptons) we  should 
get just same imaginary part of action 
contribution from a neutron and from such
a combination of protton and electron.
In an short time the contribtuion is
estimated as an integral over space of 
the Higgs field suppression. We here 
just assume by Taylor expansion in the 
presumably rather small Higgs field around
the quarks and leptons, that any effect 
will in first approximaton be linear
in the change in the Higgs field.
 Now  we find small
Yukawa-potential regions of size given 
by the inerse Higgs mass and centered 
around quark or lepton. A crucial little
problem making the estimation a bit 
less trivial and bit less accurate is, 
that these regions of significant 
Yukawa-potentials are {\em Lorentz 
contracted}, because of the non-zero 
velocity of say the quark it surrounds.
(The elctrons most copiously found 
in our universe have actualy very small 
velocities compared to the light velocity,
so for them Lorentz contraction is not 
important.)   

The following the reader should have in 
mind 
in order to estimate the 
contribution to the imaginary part 
of the action $S_I[history]$ under the 
assumtion of the dominant Higgs mass term
for a netron relative a pair of proton 
and an electron:
\begin{itemize}
\item{a} Of course - unless a linear term 
should be lacking - the contribution 
must go linearly with the Yukawa coupling 
for the quark or lepton in question.
Really the suppression of the Higgs field
arround a particle - quark or lepton say -
must go proportionally to the Higgs 
Yukawa coupling ( for fixed velocity)
\item{b} But it will vary with velocity 
due to the Lorentz contraction of the 
Higgs-Yukawa effective extendssion 
volume, around the particle.
\item{c} So at the end the effect on the
imaginary action $S_I[history]$ becomes
proportional to 
\begin{equation}
\Delta S_I[history] \propto g_{particle}
*\frac{m}{E}|_{averaged} \propto \frac{m^2
<\gamma> <\gamma^{-1}>}{E_{average}}
\end{equation}  
where $m$ is the mass of the quark say 
(or lepton) and $E$ its actual kinetic 
energy including the Einstein enrgy.
The average as the quark flies around in 
the nucleon say is denoted of its 
$\gamma=E/m$ is denoted  
$<\gamma>$, while the average of the 
inverse of this same $\gamma$ is denoted$
<\gamma^{-1}>$. The average kinetic 
including Einstein energy $E$ is denoted
$E_{average}$. The combination 
$<\gamma><\gamma^{-1}>$ would in the case 
of no fluctuations of the actual velocity 
of the quark be just unity, and thus we
may hope that we can estimate this 
product somewhat more accurately than say
its two factors separately.   
\end{itemize}

The various types of quarks have of 
course  the 
deeper Higgs
fields around them the stronger their 
Higgs Yukawa
couplings $g_{particle}$.
The Higgs field is effectively extended 
over a range of
size given by the Higgs mass but not 
dependent on the
species of quark or lepton in question.
The extend of the Yukawa potential 
rather is over an
elliptic region, that is the Lorentz 
contraction of the spherical Yukawa 
potential, which is obtained around a 
resting particle. 
So the contribution to the integral of 
the Higgs field or
presumably also over its square over all 
space from a
quark or lepton is proportional to 
$g_{particle}$ and to the
inverse of $E/m$ where $E$ is the energy 
and $m$ the mass of
the quark or lepton. The Lorentz 
contraction factor is
for Yukawa potentials for quarks
 due to motion
inside nucleons, if we have - as is 
most copiously the case - resting nucleons.
Well, really the speed of the nucleons 
inside 
the nuclei is not so negligible again but
compared to the speed of quarks inside 
nucleons it is small.

Does it Pay for ``God’’ to make Only
Neutrons or No neutrons ?

The bigger integrated Yukawa potentials 
around the quarks
and leptons the more the Higgs field is 
suppressed.
The strength of the suppressions is 
proportional to the
Yukawa coupling for particle making the 
suppression.
The extension is roughly like the 
Lorentz contracted of a sphere forming 
an ellipsoid
given by the Higgs mass(as inverse radius 
of the sphere).

The proton is almost identical to the 
neutron except, that
one up-quark has been replaced one 
down-quark.

To keep Universe chargeless a proton 
should be
accompagnied by an electron.

A neutrino typically runs so fast that 
its Yukawa potential is
much less extended in volume than those 
of quarks and charged
leptons.

\subsection{Contributions to See 
Whether
Neutrons or Non-neutrons Favored
My Prediction from Future Influence}
To estimate the contributions comming from 
a neutron to compared it to that comming 
from what is its decay products a proton 
and an electron and even a not so 
significant elctron anti neutrino we 
need the light quark masses which are not
 so well determined (and that makes our
uncertainty rather large), but let us take 
\begin{eqnarray}
m_u & = & 1.7  \hbox{to} 3.3 MeV\\
m_d & = & 4.1 \hbox{to} 5.8 MeV
\label{qmasses}
\end{eqnarray}
for respectively the up and the down 
quark masses.

In \cite{Spaatind} one arrives as also 
sketched here  
to the relation 
\begin{equation}
\sqrt{m_d^2 - m_u^2} =\sqrt{E_q m_e}
/``ln''\label{relationf}
\end{equation}
where we have denoted 
\begin{equation}
``ln'' = <\gamma><\gamma^{-1}>
\end{equation}
because this quantity for light quarks 
compared to the energy $E_{average}$ tends
to be approximately a logarithm. 
The relation(\ref{relationf})   is  
relativly well satisfied, if we take 
the quark masses 
(\ref{qmasses}), $E_q \approx 160 \ MeV$ 
and $``ln'' =2.3_7 $.(see my previous 
article\cite{} for this crude estimate) 
In fact then we would get (using 
$m_e = 0.511\  MeV$) 
\begin{eqnarray}
 R.H.S.& = &\sqrt{E_qm_e}/``ln'' 
= 3.81 \ MeV \\
L.H.S.& =& \sqrt{m_d^2 - m_u^2} 
= \sqrt{13._9} \hbox{to} \sqrt{22._75} 
\ MeV\\
& = &
3.7_3 \hbox{to} 4.7_7 \ MeV. 
\end{eqnarray}


\section{Fine Tuning Calls for 
Influence 
Going Back in Time} 
One argument, which Don Bennett and myself 
would give for some influence from the 
future being called for, is this:

We know the fine tuning problem of 
why the cosmological constant/dark energy 
/energy density in the vacuum is so small
compared to the {\em energy density} given 
by the most fundamental constants $G$, $c$,
and $\bar{h}$, i.e. the Planck energy 
density? The ratio of the actual vacuum
energy density to the from the dimensional 
arguments expected value is enormously 
small. So it is clear that there must 
have been some enormous fine tuning 
arranging this enormously small energy
density in the vacuum. Now we expect that 
the vacuum energy density should be 
constant as time has gone on. So even in 
a time of say minutes after the start 
of the universe or Big bang or whatever
the vacuum energy should have had the 
present extrememly small value. But now 
at these early times there were so big 
energy densities of radiation or matter 
that the present small vacuum energy 
density would be very small and 
insignificant compared to radiation energy
density. But when it were at that time 
so insignificant, how could {\em at that 
time} any physical effect have made a so
precisely close to zero as the vacuum 
energy density to day? So it seems that 
an influence from the future somehow must 
have arranged at this early stage already 
the exceedingly small energy density in 
vacuum? It is of course because of 
an argument in the direction of this that
is the reason for that, when Weinberg 
looks through the various explanations 
for the cosmological constant being so 
small, then the most promissing 
explanation is to use antropic principle.
The entropic principle, which states that 
parameters shall be so arranged that 
humans can come to exist, is namely 
in reality a method to to arrange a 
simulated effect of the future influencing 
the past. By thorowing away the 
scenarious which happen not to allow for 
humans one has got what functions as 
a back in time effect.
        
\section{Our Multiple Point 
Principle}
There is one very general deduction from 
such a theory with a principle of 
minmizing some quantity as we above told 
that the imaginary part $S_I[history]$
would be minimized for the actually 
realized history. This deduction would
be best achieved if we instead of 
minmizing over histories of the uiverse
minmized over combinations/sets  of 
coupling constants, but since one could 
imagine some vacuum being selcted among
several at least the effective coupling 
constants relevant for the by 
a quantity like $S_I[histoty]$ selected 
vacuum would effectively have been 
determined as if they were adjusted to 
minimize something ($S_I$) by adjusting 
the coupling constant combination. 
The deduction related to is found an 
article by Ninomiya and myself 
\cite{deductionMPP} in the Bled 
proceedings from .... The point is,
however, to imagine that  the right 
combination of coupling constants is 
achieved by asking to obtain the minimum
for some quantity - in fact our 
$S_I$, which we now imagine to depend 
also on the coupling constants
( with an effecive vacuum providing such 
couplings this imagination would be true
in our model) - under the restriction that
the energy density of the various (local)
ground states the vacua should be positve.
This assumption of vacuum energy density
being positive may be understandable in 
our model - as well as phenomenologically 
suppoted as a principle - by noting, that
if a vacuum gets (appreciably) less energy
density than zero, then the usual 
vacuum becomes unstable againts making
a transition to this low energy density 
vacuum. From the point of view of the 
history being selcted such an instability
would mean that it would be this vacuum 
rather than the usual one that got 
realized and the potential history
meant as a history in the ``usual''
vacuum would no longer be realized; so 
if this latter history gave a smallest 
$S_I$ that would be a lost achievement
if another vacuum tales over. So one 
should avoid competing vacuum threadening
the stability severely for the realized 
one, or one should presumably preferably
think that there are several vacua getting 
their realization in a turn adjuted to 
be the most beneficial for the $S_I$ 
being as negative as possible. Also in 
such a scenario of several vacua comming 
to exist as time passes on, the transition 
from to the next should not be too quick,
they are to exist for of the order of 13 
milliard years. Thus they should be 
approximately stable and we would obtain 
an approximate multiple point principle
in such a scenario. In any case we already
earlier argued that once you have the 
minimization of something like our $S_I$
that just can mannage some way to 
effectively depend on the coupling 
constants, then the couplings get 
very likely adjusted to lead to several
degenerate vacua, meaning multiple point
principle.

Having in mind that this multiple point
principle is thus to be considered a 
deduction from a minimization of some 
quantity model including future in such 
a way that it really means influence from
the future, we can now look at successes 
of our multiple point point principle 
(MPP) as also being evedence for there 
existing in the laws og nature some 
influence from the future. 

Now I remind the reader that the most 
impressive confirmation of our multiple 
point principle were that we - Colin D. Froggatt and myself - PREdicted the Higgs 
mass\cite{Higgsprediction} many years before the Higgs boson 
were found to 135 GeV $\pm$ 10 GeV !
With the present calculations and top-mass
measured our prediction would rather have
been 129.4 GeV with an uncertainty now 
rather down to about $\pm$ 1 GeV. So 
although our prediction is now only 3.4 
GeV above the experimental Higgs mass 
126 GeV, the deviation compared to the 
uncertainty may have gone slightly up 
compared to the ld day PREdiction, but
we should still consider it a great 
success for the multiple point principle
that the Higgs mass is so close to our 
prediction!          

Historically we - Don Bennett and myself 
and also in some papers with Colin D. 
Froggatt - we looked for some way of 
justifying to fit fine structure constants
by phase transition couplings in lattice 
Yang Mills theories. We worked at that 
time with what we call Anti-GUT (meaning 
anti-grand-unification) meaning that we 
rather than as were most popular to look 
for simple groups like SU(5) or SO(10) 
etc. we did not unify in the sense that 
we used the not at all simple group
$S(U(2)\times U(3))\times \cdots \times 
S(U(2)\times U(3))$ (with $N_{gen}$ cross
product factors), 
rather meaning that we gave every family 
of fermions in the Standard Model its
own family of also  gauge bosons, so that 
our ``anti grand unifying group'' were 
the cross product of one Standard Model 
gauge group, one for each family (the 
number $N_{gen}$ of families not 
yet known at that time; we had to fit it 
to the fine structure constants and 
PREdict it; luckily we PREdicted 
$N_{gen} =3$). But the problem for 
which we needed the multiple point 
principle were to give an explanation or 
at least formulate a principle that could 
imply that the {\em phase transition }
finestructure constant values were the 
ones for which Nature should care. 
But if we somehow had derived that 
Nature should have a couple (or more)
energy density wise degenerate 
vacua/phases of course if nature really 
were a lattice Yang Mills theory, then 
it would mean that Nature should choose
the phase transition value of the 
coupling constant/the finestructure 
constant.

Once we have suggested to believe in such 
a multiple point principle in the form of
there being many/several energy-wise
degenerate vacua, you just have to find 
one with an appropriate  small 
cosmological constant and you can so to 
speak transfer that small energy density 
to other vacua, thus explaining the 
smallness and even fit {\em the} 
cosmological constant (or the dark energy).
Roman Nevzorov Fraggatt and me did such 
an application in several versions, 
explaining the cosmological 
constant\cite{Roman}.               

We even mannaged to make a solution 
of the scale problem (related to the 
hierarchy problem) in the sense of using 
the postulation of the multiple point 
principle to fix the scale of the weak 
interactions (compared to the Planck 
scale, taken as the fundamental scale).
This we -Colin D.  Froggatt, Larisa 
Laperashvili and myself - did by 
speculating up the existence of a further 
vacuum in which there is a Boson 
condensate of bound states of 6 top 
and 6 anti-top quarks. In the spirit 
of the multiple point principle 
postulating a further vacuum is somewhat 
natural, and at least each time we 
postulate a new vacuum, we get the 
information out of multiple point 
principle that this vacuum shall have 
the same energy density as the other 
vacua. Thus for each new vacuum we 
postulate - and take to be degenrate 
with the other ones - we get one more of 
the say Standard Model (if that is what 
we use) determined, because one more 
relation among them is obtained. Luckily
it turns out that we essentially may 
use this new 
information to fix the weak energy scale
and most importantly:

{\em We get the weak scale out as 
restriction on between which values of the
running top-Yukawa coupling $g_t(\mu)$ 
shall be taken on at 1) the high field
scale of the second Higgs field effective
potential minimum (assumed by us to be 
essentially the Planck scale) and 2)
the weak scale.}   

Since then the running top Yukawa 
coupling must ``run'' between the 
two predicted values $g_t(\mu = 18^{18}
GeV)$ and 
$g_t(\mu = \hbox{``weak scale''})=1.02$,
the ratio of the weak to the supposed 
more fundamental scale gets predicted
to be ``exponentially'' small! Really
the point is that with the rather weak
couplings of the Standard Model the 
`running'' is actually a bit slow as 
a function of the logarithm of 
$\mu$. Thius to get a given distance 
of change in the Yukawa coupling an 
exponentially big ratio of scales is
needed. Actually our prediction of the 
logarithm of the scale ratio, the 
scale problem gets very well!

So our multiple point principle is here 
a great success: both explaining the 
exponential smallness and giving a good 
value for its logarithm.

\section{Do we have Enough 
Evidence for Influence from 
Future?}
I would like towards the end very 
optimistically for the hypotesis of there 
being indeed an influece form the future 
to give - the relatively optimistic, but
still crudely true - numbers for how 
unlikely it would be that our small 
coincidences favouring the complex action 
model with the asssumption that the Higgs 
field square dominates without such 
a model being ture. 

Say we look at the coincidence that the 
``knee'' in cosmic radiation spectrum 
just order of magnitudewise happens to
coincide with the threshold for Higgs
production. If we say one has studied 
cosmic rays from some electron volts 
up to say $10^{20}$ electron volts, we 
could say over 19 orders of magnitude. 
Then if one finds a knee to coincide within one or two orders of magnitude, it 
represents a coincidence that should 
happen by accident only in about 1/10 
cases. Similarly looking at the agreement
of our formula (\ref{relationf}) as being
that we get inside the right interval 
of length one $MeV$ for quantities 
- sides of the equation - being of order 
of 4 $MeV$, this is omething that should
only happen in one out of four cases.  

Our argument that the Higgs-field vacuum 
expecctation value should just have 
gotten that value, that minimizes the 
{\em squared} Higgs field expectaion 
value - we get agreement up to some 
factor of one or two orders of magnitude -
means that our minimization principle
led to the right order of magnitude for 
the weak/Higgs field scale to say a couple of orders of magniutde out of 17 orders 
of magnitude (taking the Planck scale 
as the fundamental one). this means again
that our influence from future got the 
right scale among say 17/2 $ \approx$ 10.

We may even count here the smallness of the binding energy in nuclei compared to the 
separately bigger kinetic and potential 
energy of the nucleons, say one out of 
2 cases accident. 

These ``numerical'' coincidences together
would give us a one out of 800 coincidence,
which is a factor 4 more than 3 standard deviations. Taking this optimistic estimate
seriously we really have more than 3 
standard deviation evedence for the
influence from future seeking to minimize
the Higgs field square (integrated over
space time), so as to use it to tune 
some couplings or the like.       

Further to support this complex action 
with Higgs mass (square) term dominating 
model for the  development of 
the world being supported we should 
collect also the evindence comming 
from the very bad lick of the S.S.C.
machine, that would if it had worked 
according to plans have produced more 
Higgs bosons than L.H.C. has so far, and 
the -  for our model though too little -
bad luck of an explosion in the tunnel, 
which though were repaired and mainly so
far had the effect of making the 
physicists choose to postpone the 
running of the L.H.C. with its planned 
beam energy of 7 TeV against 7 TeV
(meaning $\sqrt{s}=14$ TeV) till 2015.
Although it now looks that finally it 
will come to run, we may though consider 
it, that this caused postponing of the 
full energy could be a result of our 
complex action model with Higgs mass term
dominance. Together we might consider 
these after all not so terribly 
miraculous bad lucks for Higgs producing
machines as something that would not 
be at least the very first expectation 
without theory predicting it like ours.
So we might say e.g. that in at most 
one out of say 5 cases would so much 
bad luck hit the Higgs producing machines.

If we ccombine this estimate with the 
just counted, we would say that now the 
Higgs mass square term dominated complex 
action model has scored a success corresponding to one out of 800 *5 = 4000 cases!

If we add to this counting the evidence 
comming from say the Higgs mass being 
PREdicted from our multiple point 
principle, which also would follow from
an influence from the future type theory,
and take it that the range for Higgs mass
were at first up to 600 GeV or just use 
the actually Higgs mass to set the scale 
for Higgs masses the deviation $129.4
- 126$ GeV = 3.4 GeV (relative to 
respectively 600GeV  or 126 GeV) means 
a luck for our multiple point principle 
as one out of $\approx 200$ or 
one out of $\approx 36$ respectively.

If we already have counted the luck of 
our theory of getting the right weak 
scale it might no longer be new 
prediction to use the multiple point 
principle to predict the top -Yukawa
coupling to be $1.02\pm 14\%$ (oterwise
this result should give a one out of 7
good luck for our model).

Also it would probably be too much to 
seek to include as a result of our 
influence the very remarkable smallness 
of the cosmological constant because 
this influence from future type theory 
in itself does not predict this smallness,
although firstly it is very hard to see 
how such a small cosmological costant 
could come without an influence from the 
future and secondly we have works with 
Roman Nevzorov et al. \cite{Roman} in 
which we actually even fit the 
cosmological well using the multiple point
principle(which indeed is consequence 
of an influence from the future much 
like the one we discuss here. If we 
include this cosmological constant 
as were it prediction it would increase 
much our measure of the success since even
counted only as a success on the 
logarithmic we could a priori have expect 
a ``Planck energy density value'' about 
100 orders of magnitude larger. Counting 
with natural logarithm say we should then 
say we succeeded as one ot of 100*2.3 
= 230.           

But even as presumably most fair leaving 
out the cosmological constant proper as 
being a success for our model(s), but only 
taking in the Higgs mass PREdisction from 
multiple point principle (after replacing 
the one prediction of MPP by the Higgs or
weak scale gotten by adjusting this scale
to minimize the squared Higgs field 
integrated over space and time) we get 
that good luck for our model is of the 
order of getting one out of $4000 * 7
\approx 30000$ cases/possibilities correct!

This of course were optimistically counted,
but it sounds that one should take 
possibility of there being effects from 
the future. especially we did not even in 
this number include anything from the 
arguments related to the need for
ultravioet cut off, which especially for 
gravit may be very hard without a bit 
of non-locallity, thereby allowing the influence from the future sneak in in 
principle.

\section{Conclusion and Outlook}
We have in the present article looked at a
series of arguments for that there should 
be in the laws of nature some law that 
makes e.g. the initial conditions or 
the coupling constants or both be adjusted 
as if it were with a special purpose (such
as as here suggested to make a certain quantity depending on the history ``the 
imaginary part of the action be minimal).

The main classes of arguments, which 
I suggested are:

\begin{itemize}
\item Numerical or observational successes of assumtions
involving such an influence from the 
future. This includes:
\begin{itemize}
\item The bad luck of SSC, and if we 
take it seriously the very minute bad 
luck of the  LHC, both machines 
(potentially) producing relative to human
history exceptionally many Higgs bosons.
\item Our relation relating the light 
quak mass square difference to the 
electron mass square and the fraction of 
energy carried by the quarks in the 
nucleons. This relation just organizes 
that the contribution from a neutron 
and from an electron and a proton 
(and an electron anti neutrino) together
to this imaginary action is same. Thus
when this relation - which seems to be 
fullfilled within errrors in nature - 
happens to be fullfilled there would be 
no gain in minimizing the imaginary 
part of the action by neither arranging 
for more neutrons than for more of its 
decay products electron + proton (+
anti electron neutrino). The world would 
potentially be able to exist at a minimum 
for the imaginary part of the action.

\item analogously I argued that including 
the effects of virtual top quarks in the 
vacuum it could within errors be so that 
the Higgs field {\em square} is in fact
at a minimum with just the present Higgs 
expectation value in the vacuum. So indeed 
the parameters of the Standard Model 
could have been arranged just so as
to minimize the Higgs field square, and
that could have led just to the from 
hierarchy problem consideration rather 
difficult to accept compared to the 
Planck scale or Grand Unification scale
point of view exceptionally small value 
Higgs field expectation value.
\item Even the ``knee'' in the cosmic 
ray spectrum is so close to the threshold
for the severe production of Higgs
bosons that we can claim that it is as if
it had been arranged to be just like that
to make the production of Higgs bosons
by the cosmic rays hidding material 
or planets etc. in the gallaxes so small
as possible under some restrictions. 
  
The ``God'' did not quite switch off the 
cosmic rays above the effective Higgs 
production threshold, but the ``knee'' 
looks like a weak attempt to do so.

\end{itemize} 
\item We called attention to that cut off 
methods which are needed to make 
especially renormalizable gravity 
theories are very hard if at all possible
to conceive of without some non-locallity.
And then since non-locallity really means
that influence from future is getting 
allowed for small distances, also such 
cut off needs in fact calls strongly for
that influence from future cannot be 
totally avoided. We looked especially as
an example on string theory in the recent 
formulation of Ninomiya and myself. In 
this model the for the cut off effectivity
crucial feature - the ``stringyness'' one 
could say - is put in as an initial state
- and even as a {\em final state } - 
condition! If instead of or in addition
to inclusion of gravity you also want 
to have more than the experimental 
number of dimensions 3+1, the need for 
such 
cut offs that in turn leads to 
non-locallity and thereby formally admits 
influence from future gets even stronger.
\item We also mentioned the old worries 
about that the usually assumed laws of 
nature for the initial conditions and 
those for the equations of motion 
do not have the same CPT ot say just 
time reveral invariance: The initial 
conditions usually assumed are only 
for the {\em initial state}, but not 
for the final state also as a timereversal
invariant theory would have to have it.
So again some influence from future is
called for in order to make the symmetry 
be a least formally uphold.
\item Although I did not go so deeply 
into it in the present article, it is of course also one of the arguments for 
influence future comming into the 
physical theory that one in general 
relativity has wormholes and baby 
universesetc. very easily leading to 
time machines. Such time machines namely 
leads to inconsistencies unless the hapenings are finetuned to just make things go 
in a with the time machine consitent 
manner. This has been discussed by Novikov
\cite{Novikov}.

\end{itemize} 

We will at the end stress that with the 
lists of arguments in the present article
one should at least admit that the 
absolutely safe belief that there is no 
influence from the future deserves being 
investigated and confronted with as much 
knowledge as we can collect concerning 
this question. If one truly will 
uphold this absolutely safe belief that 
nothing from the future can influence 
us in any way, there is really no 
government of the universe - at least 
no government with any interest in the 
future fundamentally - then one would have
to throw away as bad 
science/misunderstandings or pure (poetic)
invention all stories about the 
government of God or destinies or the like
which may be found in mythology in the 
holy texts or the like.

At least I hope to have put a little 
doubt on the valididty of this by now 
in first approximation well working 
law of nature that future cannot influence
anything in past or now and that there is
no government of the universe whatsoever.

Instead one could look at it that the 
strong belief in this no influence from 
future nor
government arranging for the future
will turn out to be only something 
humanity believed in a relatively short
historical era from Darwin Wallace Lamark
to some day may be next year when a truly
bad luck for LHC e.g. would convince 
humanity that there exists a ``God''
(here in quotation marks) that hate the 
Higgs sufficiently to stop 
a Higgs producing machine before it gets 
produced too many Higgses!      

\section*{Acknowledgements}
One of us (HBN) thanks the Niels Bohr Institute for allowing him to stay as emeritus,
and Matias Breskvar for economic support 
to visit the Bled Conference. 


\end{document}